\documentclass[prd,aps,showpacs,preprintnumbers,amssymb]{revtex4}
\usepackage{axodraw}
\usepackage{color}
\usepackage{epsf}

\def\e3p{$\eta \rightarrow 3 \pi$}

\begin{document}
\title{%
\hfill{\normalsize\vbox{%
\hbox{}
 }}\\
{A nonperturbative method for the Yang Mills Lagrangian}}

\author{Renata Jora
$^{\it \bf a}$~\footnote[2]{Email:
 rjora@theory.nipne.ro}}

\affiliation{$^{\bf \it b}$ National Institute of Physics and Nuclear Engineering PO Box MG-6, Bucharest-Magurele, Romania}

\date{\today}

\begin{abstract}
Using the properties of the partition function for a Yang Mills theory we compute simple relations among the renormalization constants. In the particular case of the background gauge field method we obtain that the all orders beta function for the gauge coupling constant contains only the first two orders coefficients different than zero and thus corresponds to the 't Hooft scheme.
\end{abstract}
\maketitle

\section{Introduction}

One of the most difficult tasks in  particle physics calculations is to compute higher order corrections to  cross sections, decay rates or simply quantum correlators. Up to now the beta function for QED has been computed up to the fifth order whereas for QCD up to the fourth one \cite{Vladimirov}-\cite{Baikov}.
For the standard model the beta function for the three gauge couplings has been computed in the minimal subtraction scheme at three loops \cite{Mihaila}. However there is a renormalization procedure \cite{Hooft1}, \cite{Hooft2}, the 't Hooft scheme, where the beta function of the gauge coupling constant stops at two loop.

In \cite{Jora} we introduced nonperturbative methods for computing an all order correction to the mass of the scalar in the $\Phi^4$ theory  whereas in \cite{Jora1} we computed  in a semiperturbative method the beta function for QED with fermions in the fundamental representation. Unfortunately although the methods were based on the same idea for each case we were forced to introduce new techniques thus making the approach difficult to use for a general case.  In this paper we introduce a new method for the Yang Mills theory which can be applied for any renormalizable field theory  directly. Instead of relying on a perturbative approach our purpose was to determine global properties of the theory from the specific properties of the zero current partition function. We rely on the path integral formalism to obtain useful relations between the renormalization constants of the Yang Mills theory. These relations lead for the background gauge field method to a derivation of the general form of the beta function. It turns out that through this method one obtains an all loop beta function with only  the first two coefficients different than zero thus indicating that the 't Hooft scheme might be in a sense the most natural scheme for beta functions.

\section{Yang Mills partition function}
We start with the Yang Mills Lagrangian:
\begin{eqnarray}
{\cal L}=-\frac{1}{4}(F^a_{\mu\nu})^2,
\label{rez5467}
\end{eqnarray}
where,
\begin{eqnarray}
F^a_{\mu\nu}=\partial_{\mu}A^a_{\nu}-\partial_{\nu}A^a_{\mu}+gf^{abc}A^b_{\mu}A^c_{\nu}.
\label{ten45678}
\end{eqnarray}

The Lagrangian in Eq. (\ref{rez5467}) needs fixing. This is done by introducing the ghost Lagrangian:
\begin{eqnarray}
{\cal L}_{g}=\bar{c}^a(-\partial^{\mu}\partial_{\mu}-g f^{abc}\partial^{\mu}A^b_{\mu})c^c.
\label{ghost567}
\end{eqnarray}
We shall work in the Feynman gauge ($\xi=1$) and in the Fourier space throughout this paper. Thus starting from,
\begin{eqnarray}
A^a_{\mu}(x)=\frac{1}{V}\sum_n\exp[-i k_n x]A^a_{\mu}(k_n)
\label{fourier456}
\end{eqnarray}
we rewrite the full Lagrangian:
\begin{eqnarray}
&&\int d^4x {\cal L}=-\frac{1}{2}\frac{1}{V} \sum_n k_n^2A^{a\nu}(k_n)A^a_{\nu}(-k_n)+\frac{1}{V}\sum_n k_n^2 \bar{c}^a(k_n)c^a(-k_n)+
\nonumber\\
&&+\frac{i}{V^2}g\sum_{n,m} k_n^{\mu}A^a_{\nu}(k_n)f^{abc}A^b_{\mu}(k_m)A^{c\nu}(-k_n-k_m)-
\nonumber\\
&&-\frac{1}{V^3}g^2f^{abc}f^{ade}\sum_{n,m,p}A^{b\mu}(k_n)A^{c\nu}(k_m)A^d_{\mu}(k_p)A^e_{\nu}(-k_n-k_m-k_p)-
\nonumber\\
&&-\frac{i}{V^2}\sum_{n,m}k_n^{\mu}\bar{c}^a(k_n)gf^{abc}A^b_{\mu}(k_m)c^c(-k_n-k_m).
\label{four65788}
\end{eqnarray}

Then one defines the zero current partition function by the expression:
\begin{eqnarray}
Z_0=\int \prod_i\prod_j\prod_m d A^a_{\mu}(k_i) d \bar{c}^b(k_j)d c^d(k_m)\exp[i\int d^4x {\cal L}],
\label{part456}
\end{eqnarray}
where in the exponent one should use the Eq. (\ref{four65788}).

It is useful at this stage to settle some of the properties of $Z_0$. It is known that $Z_0$ apart from a factor in front is given by the exponential of the sum of all disconnected diagrams:
\begin{eqnarray}
Z_0={\rm factor}\times \exp[\sum_i V_i]
\label{disc4567}
\end{eqnarray}
where $V_i$ is a typical disconnected diagram. Since the calculation is done in the absence of external sources all $V_i$ diagrams are closed and contain summations over momenta (that appear in propagators or vertices) and thus do not depend at all on any momenta. The factor in front is a product obtained form integrating the gaussian integrals corresponding to the kinetic terms.
So one can write:
\begin{eqnarray}
Z_0= {\rm const} \prod_i (k_i^2)^{N^2-1}\prod_j (k_j^2)^{-d/2(N^2-1)}\exp[\sum_i V_i]
\label{part4567}
\end{eqnarray}
where N is coming from the Yang Mills group $SU(N)$ and the first factor corresponds to the ghosts whereas the second to the gluon fields.

We write:
\begin{eqnarray}
&&Z_0=\int \prod_i \prod_j \prod_m d A^a_{\mu}(k_i) d \bar{c}^b(k_j) d c^d (k_m) \exp[i \int d^4x{\cal L}]=
\nonumber\\
&&=\int \prod_i \prod_j \prod_m d A^a_{\mu}(k_i) d \bar{c}^b(k_j) d c^d (k_m) \frac{ d A^a_{\nu}(k)}{d A^a_{\nu}(k)}\exp[i \int d^4x{\cal L}]=
\nonumber\\
&&= \int \prod_i \prod_j \prod_m d A^a_{\mu}(k_i) d \bar{c}^b(k_j) d c^d (k_m) \frac{d}{d A^a_{\nu}(k)}[ A^a_{\nu}(k)\exp[i \int d^4x{\cal L}]]-
\nonumber\\
&&-\int \prod_i \prod_j \prod_m d A^a_{\mu}(k_i) d \bar{c}^b(k_j) d c^d (k_m) A^a_{\nu}(k)\frac{d}{d A^a_{\nu}(k)}\exp[i \int d^4x{\cal L}].
\label{rez54678}
\end{eqnarray}
We start by analyzing the first term on the right side of the Eq. (\ref{rez54678}) to get:
\begin{eqnarray}
&&\int \prod_i \prod_j \prod_m d A^a_{\mu}(k_i) d \bar{c}^b(k_j) d c^d (k_m) A^a_{\nu}(k)\exp[i \int d^4x{\cal L}]_{A^a_{\nu}(k)=+\infty}-
\nonumber\\
&&\int \prod_i \prod_j \prod_m d A^a_{\mu}(k_i) d \bar{c}^b(k_j) d c^d (k_m) A^a_{\nu}(k)\exp[i \int d^4x{\cal L}]_{A^a_{\nu}(k)=-\infty}.
\label{expr7689}
\end{eqnarray}
Although the Fourier transform of the gauge field has a real and a imaginary part we can assume that it is real without any loss of generality as the same arguments apply. We first note that the exponential factor in Eq. (\ref{expr7689}) will contain:
\begin{eqnarray}
\exp[i\int d^4x {\cal L}]\sim {\rm other\, factors}\times\exp[-\frac{i}{2}k^2A^{a\nu}(k)A^a_{\nu}(k)]
\label{factor7689}
\end{eqnarray}
However $k^2$ should actually be written as $k^2+i\epsilon$ where $\epsilon$ ensures the convergence of the gaussian integral corresponding to the term in Eq. (\ref{factor7689}). Then the limits in Eq. (\ref{expr7689}) will be zero as they contain an exponential that goes to zero as it can be seen from :
\begin{eqnarray}
\lim_{A^a_{\nu}\rightarrow\pm\infty}A^a_{\nu}(k)\exp[-\frac{i}{2}k^2A^{a\nu}(k)A^a_{\nu}(k)-\frac{\epsilon}{2}A_{a\nu}(k)A^a_{\nu}(k)]=0
\label{rez456}
\end{eqnarray}
Note that we picked a space time component $\nu$ such that the corresponding metric for it is $g^{\nu\nu}=-1$. Thus the first contribution on the right hand side of the  Eq. (\ref{rez54678}) cancels.  The second contribution is given by:
\begin{eqnarray}
&&Z_0=\int \prod_i \prod_j \prod_m d A^a_{\mu}(k_i) d \bar{c}^b(k_j) d c^d (k_m)(-i)[-\frac{k^2}{V}A^{a\nu}(k)A^a_{\nu}(-k)+
\nonumber\\
&&\frac{3i}{V^2}gk^{\mu}\sum_pf^{abc}A^a_{\nu}(k)A^b_{\mu}(p)A^{c\nu}(-k-p) -\frac{i}{V^2}g\sum_pp^{\nu} \bar{c}^b(p)f^{bac}A^{a}_{\nu}(k)c^c(-p-k)-
\nonumber\\
&&-\frac{1}{V^3}g^2f^{bac}f^{bde}\sum_{p,q}A^a_{\nu}(k)A^c_{\mu}(p)A^{d\nu}(q)A^{e\mu}(-p-k-q)]\times\exp[i\int d^4x {\cal L}].
\label{one65789}
\end{eqnarray}

According to Eq. (\ref{part4567}) one can write:
\begin{eqnarray}
k^{\mu}\frac{d Z_0} {\partial k^{\mu}}=-2(N^2-1)[\frac{d}{2}-1]Z_0
\label{some324}
\end{eqnarray}

We further apply the operator $k^{\mu}\frac{d}{d k^{\mu}}$ to Eq. (\ref{part456}) to obtain:
\begin{eqnarray}
&&k^{\mu}\frac{d Z_0}{d k^{\mu}}=\int \prod_i \prod_j \prod_m d A^a_{\mu}(k_i) d \bar{c}^b(k_j) d c^d (k_m)\times
\nonumber\\
&&i[-\frac{1}{V}k^2A^{a\nu}(k)A^a_{\nu}(-k)+\frac{2}{V}k^2\bar{c}^a(k)c^a(-k)+\frac{i}{V^2}k^{\mu}\sum_{p}A^a_{\nu}(k)f^{abc}gA^b_{\mu}(p)A^{c\nu}(-p-k)-
\nonumber\\
&&-\frac{i}{V^2}\sum_{p}k^{\mu}\bar{c}^a(k)gf^{abc}A^b_{\mu}(p)c^c(-p-k)]\times\exp[i\int d^4x {\cal L}.
\label{sec4355}
\end{eqnarray}

The next step will be to reconsider the Lagrangian from the perspective of renormalization. Thus the fields are rescaled using the standard procedure which leads to:
\begin{eqnarray}
&&{\cal L}_r=-\frac{1}{2}\frac{1}{V}Z_3 \sum_n k_n^2A^{a\nu}(k_n)A^a_{\nu}(-k_n)+\frac{1}{V}Z_2\sum_n k_n^2 \bar{c}^a(k_n)c^a(-k_n)+
\nonumber\\
&&+\frac{i}{V^2}Z_{3g}g\sum_{n,m} k_n^{\mu}A^a_{\nu}(k_n)f^{abc}A^b_{\mu}(k_m)A^{c\nu}(-k_n-k_m)-
\nonumber\\
&&-\frac{1}{V^3}Z_{4g}g^2f^{abc}f^{ade}\sum_{n,m,p}A^{b\mu}(k_n)A^{c\nu}(k_m)A^d_{\mu}(k_p)A^e_{\nu}(-k_n-k_m-k_p)-
\nonumber\\
&&-\frac{i}{V^2}Z_1'\sum_{n,m}k_n^{\mu}\bar{c}^a(k_n)gf^{abc}A^b_{\mu}(k_m)c^c(-k_n-k_m).
\label{four6522788}
\end{eqnarray}
Here the fields and the couplings should be considered the renormalized ones and the renormalization constants satisfy the Slanov-Taylor identities:
\begin{eqnarray}
g_0^2=\frac{Z_{3g}^2}{Z_3^3}g^2\mu^{\epsilon}=\frac{Z_{4g}}{Z_3^2}g^2\mu^{\epsilon}=\frac{Z_1^{\prime 2}}{Z_2^{ 2}Z_3}g^2\mu^{\epsilon},
\label{renorm6657}
\end{eqnarray}
where $d=4-\epsilon$ and $\mu$ is a parameter with dimension of mass and we shall use dimensional regularization scheme.

As an aside note that in the background gauge field method which consists in the separation of the gauge field $A^a_{\mu}$ into a  background gauge field $B^a_{\mu}$ and a quantum fluctuation $\tilde{A}^a_{\mu}$  only the background gauge field gets renormalized as the quantum fluctuations appear only inside loops and one has in this case a simple relation among the renormalization constants:
\begin{eqnarray}
&&Z_{4g}=Z_{3g}=Z_3
\nonumber\\
&&Z_2=Z_1'
\nonumber\\
&&Z_g=Z_3^{-1/2}.
\label{rel7654}
\end{eqnarray}

Now consider that instead of applying the procedure that led to the Eqs. (\ref{one65789}) and (\ref{sec4355}) to the bare Lagrangian we apply it to the renormalized one. Then Eqs.(\ref{one65789}) and (\ref{sec4355}) will become:
\begin{eqnarray}
&&Z_0=\int \prod_i \prod_j \prod_m d A^a_{\mu}(k_i) d \bar{c}^b(k_j) d c^d (k_m)(-i)[-\frac{k^2}{V}Z_3A^{a\nu}(k)A^a_{\nu}(-k)+
\nonumber\\
&&\frac{3i}{V^2}Z_{3g}gk^{\mu}\sum_pf^{abc}A^a_{\nu}(k)A^b_{\mu}(p)A^{c\nu}(-k-p) -\frac{i}{V^2}g Z_1'\sum_p p^{\nu}\bar{c}^b(p)f^{bac}A^{a}_{\nu}(k)c^c(-p-k)-
\nonumber\\
&&-\frac{1}{V^3}g^2Z_{4g}f^{bac}f^{bde}\sum_{p,q}A^a_{\nu}(k)A^c_{\mu}(p)A^{d\nu}(q)A^{e\mu}(-p-k-q)]\times\exp[i\int d^4x {\cal L}],
\label{one2365789}
\end{eqnarray}
and
\begin{eqnarray}
&&-2(N^2-1)[\frac{d}{2}-1]Z_0=\int \prod_i \prod_j \prod_m d A^a_{\mu}(k_i) d \bar{c}^b(k_j) d c^d (k_m)\times
\nonumber\\
&&i[-\frac{1}{V}Z_3k^2A^{a\nu}(k)A^a_{\nu}(-k)+\frac{2}{V}Z_2k^2\bar{c}^a(k)c^a(-k)+\frac{i}{V^2}k^{\mu}Z_{3g}\sum_{p}A^a_{\nu}(k)f^{abc}gA^b_{\mu}(p)A^{c\nu}(-p-k)-
\nonumber\\
&&-\frac{i}{V^2}Z_1'\sum_{p}k^{\mu}\bar{c}^a(k)gf^{abc}A^b_{\mu}(p)c^c(-p-k)]\times\exp[i\int d^4x {\cal L}].
\label{se21c4355}
\end{eqnarray}

\section{Relations between the renormalization constants and the beta function}

First let us review the equivalence that exist between the interaction picture in QFT and the path integral formalism. We illustrate this for the two point functions of a gauge theory theory although this is generally applicable:
\begin{eqnarray}
&&\langle\Omega | T[A^a_{\mu}(x_1)A^b_{\nu}(x_2)] |\Omega\rangle=
\nonumber\\
&&\lim_{T\rightarrow \infty(1-i\epsilon)}\frac{\langle 0|T[A^a_{\mu}(x_1)A^b_{\nu}(x_2)]\exp[-i\int^T_{-T}dt H_I(t)]|0\rangle}{\langle 0|\exp[-i\int^T-{-T}dt H_I(t)]|0 \rangle}=
\nonumber\\
&&\lim_{T\rightarrow \infty(1-i\epsilon)}\frac{\int d A^c_{\rho} d\bar{c}^d d c^e A^a_{\mu}(x_1)A^b_{\nu}(x_2)\exp[i\int d^4x {\cal L}]}{\int d A^c_{\rho}\exp[i\int d^4x {\cal L}]}
\label{equibrel65789}
\end{eqnarray}

To this we should add the known LSZ reduction formula which can be applied similarly to the interaction picture and to the path integral formalism:
\begin{eqnarray}
&&\langle \Omega|T[A^a_{\mu}(p_1)...A^d_{\nu}(p_m)A^b_{\rho}(k_1)...A^e_{\sigma}(k_n)]|\Omega\rangle\sim
\nonumber\\
&&\sim_{p_i^0(k_j^0)\rightarrow E_{\vec{p}_i}(E_{\vec{k}_j})}{\rm polarization\,factor}\times{\rm const}\times\langle \vec{p}_1...\vec{p}_m|S|\vec{k}_1...\vec{k}_n\rangle
\left(\prod_{i=1}^m\frac{i Z_3^{1/2}}{p_i^2+i\epsilon}\right) \left(\prod_{j=1}^n\frac{i Z_3^{1/2}}{k_j^2+i\epsilon}\right)
\label{LSZred546}
\end{eqnarray}

We shall apply the LSZ reduction formula following the equivalence given in Eq. (\ref{equibrel65789}) to Eqs. (\ref{one2365789}) and (\ref{se21c4355}) to get:
\begin{eqnarray}
&&1=a_1Z_3+a_2gZ_{3g}\sum_p k^{\mu}\frac{1}{k^2p^2(p+k)^2}f^{abc}\langle \vec{k},\epsilon_{a,\nu};\vec{p},\epsilon_{b,\mu}|S|\overrightarrow{p+k},\epsilon_{c,\nu}\rangle+
\nonumber\\
&&+a_3gZ_1'\sum_p p^{\mu}\frac{1}{p^2k^2(p+k)^2}f^{abc}\langle \vec{p},a;\vec{k},\epsilon_{b,\mu}|S|\overrightarrow{p+k},c\rangle+
\nonumber\\
&&+a_4g^2Z_{4g}\sum_{p,q}\frac{1}{k^2p^2q^2(k+p+q)^2}f^{bac}f^{bde}\langle \vec{k},\epsilon_{a,\nu};\vec{p},\epsilon_{c,\mu}|S|-\vec{q},\epsilon_{d,\nu};\overrightarrow{k+p+q},\epsilon_{e,\mu}\rangle,
\label{form6655}
\end{eqnarray}
and,
\begin{eqnarray}
&&-2(N^2-1)(\frac{d}{2}-1)=
b_1Z_3+b_2Z_2'+b_3Z_{3g}g\sum_p k^{\mu}\frac{1}{k^2p^2(p+k)^2}f^{abc}\langle \vec{k},\epsilon_{a,\nu};\vec{p},\epsilon_{b,\mu}|S|\overrightarrow{p+k},\epsilon_{c,\nu}\rangle+
\nonumber\\
&&b_4gZ_1'\sum_p k^{\mu}\frac{1}{p^2k^2(p+k)^2}f^{abc}\langle \vec{k},a;\vec{p},\epsilon_{b,\mu}|S|\overrightarrow{p+k},c\rangle.
\label{sec231212}
\end{eqnarray}
First note that in the brackets of the Eqs. (\ref{form6655}) and (\ref{sec231212}) appear the three point and four points vertex functions: second we did not introduce in the standard LSZ formulas the renormalization constant as we start with the renormalized Lagrangian and thus the corresponding propagators are those fixed by the renormalization conditions.

Eqs. (\ref{form6655}) and (\ref{sec231212}) contain relations between the renormalization constants and the vertex functions. From these one can derive other useful relations. We shall analyze in detail what can one deduce from the first of them Eq. (\ref{form6655}). First note that the propagators that appear in the LSZ reduction formula are on shell thus the right hand side of the Eq. (\ref{form6655}) are  divergent. Let us analyze a typical term,
\begin{eqnarray}
&&a_2Z_{3g}g\sum_p k^{\mu}\frac{1}{k^2p^2(p+k)^2}f^{abc}\langle \vec{k},\epsilon_{a,\nu};\vec{p},\epsilon_{b,\mu}|S|\overrightarrow{p+k},\epsilon_{c,\nu}\rangle=
\nonumber\\
&&a_2Z_{3g}g\sum_pk^2\frac{1}{k^2p^2(p+k)^2}f^{abc}f^{abc}\Gamma(p,k,-(p+k))+.....
\label{typ8790}
\end{eqnarray}
Here $\Gamma(p,k,-(p+k))$ represents the vertex function from which we extracted the Lorentz and internal group dependence. Note that there are other terms on the right hand side of Eq. (\ref{typ8790}) which we ignore for reasons that will be evident soon. However since we work in the Feynman gauge and all momenta $k,p,p+k$ are on shell the corresponding vertex factor can depend only on $p^2$, $k^2$, $pk$ which are zero so the corresponding vertex function can be assimilated with $\Gamma(0,0,0)$ which  by the renormalization condition is simply
$\Gamma(0,0,0)=g$. Note that the same argument would not apply for the case when the final states would contain fermions but even in this case when could just extract the convenient contribution from it. Then one can further write:
\begin{eqnarray}
&&a_2Z_{3g}\sum_pk^2\frac{1}{k^2p^2(p+k)^2}f^{abc}f^{abc}\Gamma(p,k,-(p+k))=a_2Z_{3g}\sum_p\frac{1}{p^2(p+k)^2}{\rm const}g^2=
\nonumber\\
&&a_2Z_{3g}g^2{\rm const} \frac{(p^2)^2}{p^2(k+p)^2}=b Z_{3g}g^2
\label{rez45678}
\end{eqnarray}
 Applying the same procedure to all the terms in Eq. (\ref{form6655}) one obtains:
\begin{eqnarray}
1=aZ_2+b Z_{3g}g^2+c Z_1^{\prime 2}g^2+d Z_{4g}g^4,
\label{rel777890}
\end{eqnarray}
where the coefficients $a$, $b$, $c$, $d$ are independent of the gauge coupling constant but remain undetermined. The same arguments applied to Eq. (\ref{sec231212}) lead to:
\begin{eqnarray}
x=yZ_3+z Z_2+u Z_{3g}g^2+w Z_1^{\prime}g^2
\label{sec2221}
\end{eqnarray}

To the relation obtained in Eqs. (\ref{rel777890}) and (\ref{sec2221}) one can add another one. This is based on  writing $\frac{d\bar{c}^a(k)}{d \bar{c}^a(k)}=1$ and using the property of integration of an anticommuting variable:
\begin{eqnarray}
&&\int d\theta [A+B\theta]=B
\nonumber\\
&&\int d\theta\theta \frac{d}{d \theta}[A+B\theta]=B.
\label{ant3456}
\end{eqnarray}

Applied to the partition function $Z_0$ and to the ghost fields Eq. (\ref{ant3456}) leads to another useful relation between the renormalization constants for the ghost fields:
\begin{eqnarray}
r_1=r_2Z_2+r_3Z_1'g^2,
\label{newrel8979}
\end{eqnarray}
where $r_1$, $r_2$ and $r_3$ are constant independent of the gauge coupling constant.

\section{The beta function and discussion}

In the standard approach the Eqs. (\ref{rel777890}), (\ref{sec2221}) and (\ref{newrel8979}) are useful relations  but  are not enough for determining the beta function as we have three equations and five renormalization constants. However in the background gauge field method where the relations in Eq. (\ref{rel7654}) hold the number of renormalization constants is reduced to two and one can find important information. Thus one can extract directly the connection between the renormalization constants from the equations (\ref{rel777890}), (\ref{sec2221}) and (\ref{newrel8979}) which will be rewritten as:
\begin{eqnarray}
&&1=(f_1+f_2g^2+f_3g^4)Z_3+f_4Z_2g^2
\nonumber\\
&&1=(h_1+h_2g^2)Z_3+h_4Z_2g^2+h_3Z_2
\nonumber\\
&&1=(c_1+c_2g^2)Z_2
\label{finslrel76890}
\end{eqnarray}
where $f_i$, $h_i$ and $c_i$ are constants, some of them divergent.
From the above system one determines a formula for $Z_3$ and a consistency condition:
\begin{eqnarray}
&&Z_3=\frac{(c_1-h_3)+(c_2-h_4)g^2}{(c_1+c_2g^2)(h_1+h_2g^2)}
\nonumber\\
&&\frac{(c_1-h_3)+(c_2-h_4)g^2}{h_1+h_2g^2}=\frac{c_1+(c_2-f_4)g^2}{f_1+f_2g^2+f_3g^4}
\label{usef5455}
\end{eqnarray}

The consistency condition should be regarded as an expansion in the gauge coupling constant and leads to relations among the coefficients from which one can extract the only one that simplifies $Z_3$ which is $c_2=h_4$. Then $Z_3$ becomes:
\begin{eqnarray}
Z_3=\frac{1}{1+d_1g^2+d_2g^4},
\label{rez43567}
\end{eqnarray}
where $d_1=(c_1h_2+h_1c_2)/(c_1-h_3)$, $d_2=c_2h_2/(c_1-h_3)$ and we took $\frac{c_1-h_3}{c_1h_1}=1$ ($h_1=2$, $h_3=-1$ and $c_1=1$). This fact can be deduced from the initial equations and also from the known form of the renormalization constants.

We shall work in the dimensional regularization scheme where one can write the renormalization constant $Z_3$ as:
\begin{eqnarray}
Z_3=1+\frac{Z_3^{(1)}}{\epsilon}+\frac{Z_3^{(2)}}{\epsilon^2}+...
\label{rez65789}
\end{eqnarray}
Moreover the coefficients $d_1$ and $d_2$ are divergent and can be written as:
\begin{eqnarray}
&&d_1=\frac{d_1^{(1)}}{\epsilon}+\frac{d_1^{(2)}}{\epsilon^2}+...
\nonumber\\
&&d_2=\frac{d_2^{(1)}}{\epsilon}+\frac{d_2^{(2)}}{\epsilon^2}+...
\label{coef5467}
\end{eqnarray}

 Applying Eqs. (\ref{rez65789}) and (\ref{coef5467}) to Eq. (\ref{rez43567}) one obtains:
 \begin{eqnarray}
 Z_3^{(1)}=-d_1^{(1)}g^2-d_2^{(1)}g^4
 \label{imprt5678}
  \end{eqnarray}

In dimensional regularization in the background gauge field method (see \cite{Abbott}) the beta function is defined as:
\begin{eqnarray}
\beta=\mu^2\frac{ dg^2}{d \mu^2}=-g^4\frac{\partial Z_3^{(1)}}{\partial g^2}=g^4(d_1^{(1)}+g^2d_2^{(1)})
\label{rez111}
\end{eqnarray}
Since the two first order coefficients are universal the beta function is determined completely and thus corresponds to the 't Hooft scheme. The coefficients $d_1^{(1)}$ and $d_2^{(1)}$ are then identified with:
\begin{eqnarray}
&&d_1^{(1)}=-\frac{11}{3}N\frac{1}{(4\pi)^2}
\nonumber\\
&&d_2^{(1)}=-\frac{34}{3}N^2\frac{1}{(4\pi)^4}.
\label{co7767}
\end{eqnarray}

Note that we determined an all order shape of the beta function without computing anything that resembles a Feynman diagram simply by using global properties of the partition function of a Yang Mills theory.

\section*{Acknowledgments} \vskip -.5cm

The work of R. J. was supported by a grant of the Ministry of National Education, CNCS-UEFISCDI, project number PN-II-ID-PCE-2012-4-0078.

\end{document}